\documentclass[lettersize, journal]{IEEEtran}
\usepackage{amsmath,amsfonts,amssymb}
\usepackage{bm}
\usepackage{fontawesome}
\usepackage{makecell}
\usepackage{algorithmic}
\usepackage{algorithm}
\usepackage{array}
\usepackage[caption=false,font=scriptsize,labelfont=sf,textfont=sf]{subfig}
\usepackage{textcomp}
\usepackage{stfloats}
\usepackage{url}
\usepackage{verbatim}
\usepackage{graphicx}
\usepackage{cite}
\usepackage{threeparttable}
\newtheorem{remark}{Remark}
\begin{document}
\title{Stacked Intelligent Metasurfaces for Multiuser Downlink Beamforming in the Wave Domain}
\author{Jiancheng An,~\IEEEmembership{Member,~IEEE}, Marco Di Renzo,~\IEEEmembership{Fellow,~IEEE}, M\'erouane Debbah,~\IEEEmembership{Fellow,~IEEE},\\
H. Vincent Poor,~\IEEEmembership{Life Fellow,~IEEE}, and Chau Yuen,~\IEEEmembership{Fellow,~IEEE}
\thanks{This research was supported in part by the A*STAR (Agency for Science, Technology and Research) Singapore Grant No. M22L1b0110 and Ministry of Education MOE Tier 2 (Award number MOE-T2EP50220-0019). The work of M. Di Renzo was supported in part by the European Union through the Horizon Europe project COVER under grant agreement number 101086228, the Horizon Europe project UNITE under grant agreement number 101129618, the Horizon Europe project INSTINCT under grant agreement number 101139161, and the Horizon Europe project TWIN6G under grant agreement number 101182794, as well as by the Agence Nationale de la Recherche (ANR) through the France 2030 project ANR-PEPR Networks of the Future under grant agreement NF-YACARI 22-PEFT-0005, and by the CHIST-ERA project PASSIONATE under grant agreements CHIST-ERA-22-WAI-04 and ANR-23-CHR4-0003-01. H. V. Poor would like to acknowledge the U.S National Science Foundation under Grant ECCS-2335876. \emph{(Corresponding author: Chau Yuen.)}}
\thanks{This article was presented in part at the IEEE International Conference on Communications (ICC), Rome, Italy, 2023 \cite{arXiv_2023_An_Stacked}.}
\thanks{J. An and C. Yuen are with the School of Electrical and Electronics Engineering, Nanyang Technological University, Singapore 639798 (e-mail: jiancheng\_an@163.com; chau.yuen@ntu.edu.sg).}
\thanks{M. Di Renzo is with Universit\'e Paris-Saclay, CNRS, CentraleSup\'elec, Laboratoire des Signaux et Syst\`emes, 3 Rue Joliot-Curie, 91192 Gif-sur-Yvette, France. (e-mail: marco.di-renzo@universite-paris-saclay.fr), and with King's College London, Centre for Telecommunications Research -- Department of Engineering, WC2R 2LS London, United Kingdom (e-mail: marco.di\_renzo@kcl.ac.uk).}
\thanks{M. Debbah with KU 6G Research Center, Department of Computer and Information Engineering, Khalifa University, Abu Dhabi 127788, UAE (e-mail: merouane.debbah@ku.ac.ae) and also with CentraleSupelec, University Paris-Saclay, 91192 Gif-sur-Yvette, France.}
\thanks{H. Vincent Poor is with the Department of Electrical and Computer Engineering, Princeton University, Princeton, NJ 08544 USA (e-mail: poor@princeton.edu).}\vspace{-1cm}}
\maketitle
\begin{abstract}
Intelligent metasurface has recently emerged as a promising technology that enables the customization of wireless environments by harnessing large numbers of low-cost reconfigurable scattering elements. However, prior studies have predominantly focused on single-layer metasurfaces, which have limitations in terms of wave-domain processing capabilities due to practical hardware limitations. In contrast, this paper introduces a novel stacked intelligent metasurface (SIM) design. Specifically, we investigate the integration of SIM into the downlink of a multiuser multiple-input single-output (MISO) communication system, where an SIM, consisting of a multilayer metasurface structure, is deployed at the base station (BS) to facilitate transmit beamforming in the electromagnetic wave domain. This eliminates the need for conventional digital beamforming and high-resolution digital-to-analog converters at the BS. To this end, an optimization problem is formulated to maximize the sum rate of all user equipments by jointly optimizing the transmit power allocation at the BS and the wave-based beamforming at the SIM, subject to constraints on the transmit power budget and discrete phase shifts. Furthermore, we propose a computationally efficient algorithm for solving the formulated joint optimization problem and elaborate on the potential benefits of employing SIM in wireless networks. Numerical results are illustrated to corroborate the effectiveness of the proposed SIM-enabled wave-based beamforming design and to evaluate the performance improvement achieved by the proposed algorithm compared to various benchmark schemes. It is demonstrated that considering the same number of transmit antennas, the proposed SIM-based system achieves about 200\% improvement in terms of sum rate compared to conventional MISO systems. The code for this paper is available at \url{https://github.com/JianchengAn}.
\end{abstract}


\section{Introduction}
Over the past decade, various advanced wireless technologies such as millimeter-wave communications and massive multiple-input multiple-output (MIMO) systems have been developed to enhance network capacity and enable ubiquitous wireless connectivity \cite{CM_2014_Boccardi_Five}. However, the practical implementation of these technologies is restricted by their excessive energy consumption and costly hardware equipment \cite{JSTSP_2016_Sohrabi_Hybrid, TCOM_2015_Rappaport_Wideband, TGCN_2022_An_Joint}. Therefore, the future evolution of wireless networks needs a fundamental paradigm shift in priorities from solely increasing network capacity to focusing on energy sustainability \cite{JSAC_2020_Zhang_Prospective}. Within this context, reconfigurable intelligent surfaces (RISs) have recently emerged as a disruptive technology that can effectively improve both spectral and energy efficiency in wireless networks \cite{JSAC_2020_Renzo_Smart, CST_2021_Liu_Reconfigurable, TWC_2019_Huang_Reconfigurable, TCOM_20202_An_Low, TCOM_2021_Wan_Terahertz}. In general, an RIS is made of an artificial metasurface consisting of a large number of low-cost, nearly passive elements. Each element can independently impose an adjustable phase shift on the incident electromagnetic (EM) waves \cite{TC_2021_Wu_Intelligent, JSTSP_2022_Pan_An, WCL_2022_An_Scalable}. By adjusting the phase shifts of all the elements with the aid of a smart controller (e.g., a field programmable gate array (FPGA)), RISs are capable of manipulating the reflected and/or transmitted EM waves, thus creating favorable propagation environments by dynamically shaping wireless channels \cite{TWC_2020_Nadeem_Asymptotic, TCOM_2020_Wu_Beamforming, WCL_2021_An_The}. Additionally, RISs operate in a nearly passive mode and can be easily integrated into existing cellular/WiFi network infrastructures \cite{TC_2021_Wu_Intelligent, TC_2022_Tang_Path, CST_2021_Liu_Reconfigurable, WC_2022_An_Codebook}. Thanks to these outstanding features, RISs have attracted considerable research interest for enhancing the quality-of-service (QoS) performance of wireless networks in various typical communication scenarios \cite{JSAC_2020_Di_Hybrid, TCCN_2022_Xu_Time, JSAC_2020_Yu_Robust, TVT_2022_Xu_Reconfigurable, TWC_2020_Guo_Weighted, WCL_2022_Xu_Deep, TWC_2022_Van_Reconfigurable, TWC_2023_Xu_Antenna, TWC_2022_Papazafeiropoulos_Intelligent, TCOM_2025_An_Flexible, WCL_2023_Jia_Environment, TVT_2019_Han_Large, WCL_2025_Yu_Weighted, TWC_2025_An_Flexible}.

Nevertheless, existing research efforts on RIS-assisted wireless networks, e.g., \cite{JSTSP_2022_Pan_An, WCL_2022_An_Scalable, TWC_2020_Nadeem_Asymptotic, TCOM_2020_Wu_Beamforming, WCL_2021_An_The, TC_2021_Wu_Intelligent, TC_2022_Tang_Path, WC_2022_An_Codebook, JSAC_2020_Di_Hybrid, TCCN_2022_Xu_Time, JSAC_2020_Yu_Robust, TVT_2022_Xu_Reconfigurable, TWC_2020_Guo_Weighted, WCL_2022_Xu_Deep, TWC_2022_Van_Reconfigurable, TWC_2023_Xu_Antenna, TWC_2022_Papazafeiropoulos_Intelligent}, typically rely on a single-layer metasurface structure, which confines the available optimization variables for adjusting the beam patterns. Furthermore, for the sake of reducing the implementation cost of RIS deployments, each reflecting element is restricted to have only a discrete number of phase shifts, inevitably leading to beam misalignments in the desired service area and impairing the anticipated performance gains \cite{TCOM_2020_Wu_Beamforming, TCOM_20202_An_Low}. Additionally, recent research progress has demonstrated that RISs lack the capability of suppressing the inter-user interference, since typical single-layer implementations optimize only the phase of incident waves \cite{TWC_2020_Guo_Weighted}.

\begin{table*}[!t]
\renewcommand\arraystretch{2}
\caption{A comparison between wave-based beamforming and other beamforming schemes}
\label{tab1}
\centering
\begin{threeparttable}
\begin{tabular}{l||l|c|c|c|c|c}
\hline
Beamforming scheme & Hardware implementation & \makecell[c]{Computing\\speed}& \makecell[c]{Hardware cost}& \makecell[c]{Energy\\consumption}& \makecell[c]{ADC/DAC\\resolution} &\makecell[c]{Number of\\RF chains}\\
\hline
Fully digital \cite{TSP_2004_Spencer_Zero} & Baseband microprocessor & Slow & High& High& High & Large\\
\hline
Hybrid digital and analog \cite{JSTSP_2016_Sohrabi_Hybrid}& \makecell[l]{Baseband microprocessor,\\and analog phase shifters} & Moderate & Moderate& Moderate& Moderate & Moderate\\
\hline
Hybrid active and passive \cite{TCOM_2020_Wu_Beamforming}& \makecell[l]{Baseband microprocessor,\\and single programmable metasurface} & Fast & Low& Low& Moderate& Moderate\\
\hline
Proposed wave-based $\star$ & Multiple programmable metasurfaces& Very fast& Very low& Very low& Low& Small\\
\hline
\end{tabular}
\end{threeparttable}
\end{table*}

Motivated by these observations, a stacked intelligent metasurface (SIM) device has emerged recently \cite{JSAC_2023_An_Stacked, arXiv_2023_An_Stacked, ICASSP_2025_Lin_UAV, TCCN_2025_Liu_Multi, TWC_2025_Shi_Joint, TCOM_2025_Li_Stacked, TIFS_2025_Niu_On, TAP_2025_An_Emerging, WCL_2025_Huang_Stacked, WCL_2024_Niu_Stacked, JSAC_2024_An_Two}. By stacking an array of programmable transmissive metasurfaces, an SIM has a structure similar to artificial neural networks (ANN), thus having remarkable signal processing capabilities compared to its single-layer counterpart, e.g., an RIS. In addition, the forward propagation in SIM is executed at the speed of light, which is in contrast to conventional ANNs whose computational speed depends on existing commercial microprocessors. SIM is firmly grounded in the latest wave-based computing technology and tangible hardware prototypes rather than being an abstract concept \cite{Science_2018_Lin_All, NE_2022_Liu_A}. Specifically, Liu \emph{et al.} \cite{NE_2022_Liu_A} designed a programmable SIM, where each meta-atom acts as a reprogrammable artificial neuron. They demonstrated that this programmable SIM is capable of executing various complex signal processing and computational tasks, such as image classification, by flexibly manipulating the EM waves propagating through its multiple layers \cite{NE_2022_Liu_A}. Furthermore, an SIM-based transceiver design was proposed in \cite{JSAC_2023_An_Stacked} for point-to-point MIMO communication systems. Different from conventional MIMO designs, an SIM is able to automatically accomplish precoding and combining as the EM waves propagate through it, allowing each data stream to be radiated and recovered independently at the corresponding transmit and receive ports. When considering practical digital modulation, only low-resolution analog-to-digital converters (ADCs) and digital-to-analog converters (DACs) are required without compromising the error performance \cite{arxiv_2023_An_Stacked_mag}. This is in contrast to the utilization of low-resolution ADCs/DACs in massive MIMO systems, which typically results in a non-negligible performance penalty \cite{TCOM_2016_Choi_Near}.

In our previous work \cite{arXiv_2023_An_Stacked}, an SIM-based transceiver design for multiuser MISO downlink communication systems was developed. Therein, an SIM was integrated with the radome of the base station (BS) to perform downlink multiuser beamforming directly in the EM domain. This novel paradigm eliminates the need for digital beamforming and operates with a moderate number of radio frequency (RF) chains, which significantly reduces the hardware cost and energy consumption, while substantially decreasing the precoding delay thanks to the processing in the wave domain \cite{JSAC_2023_An_Stacked}. However, in \cite{arXiv_2023_An_Stacked}, the meta-atoms of the SIM were assumed to apply any tuning coefficient, which may be difficult to achieve in real-world applications. Against this background, this paper extends the SIM-based transceiver in \cite{arXiv_2023_An_Stacked} by considering meta-atoms that can only be tuned discretely.

Before proceeding further, we explicitly contrast the proposed SIM-enabled wave-based beamforming\footnote{The term \emph{wave-based beamforming} refers to performing the matrix multiplication required by the transmit precoding in the wave domain. This process naturally occurs as the transmitted signals propagate through the SIM.} scheme with the conventional counterparts including fully digital beamforming \cite{TSP_2004_Spencer_Zero}, hybrid digital and analog beamforming \cite{JSTSP_2016_Sohrabi_Hybrid}, as well as hybrid active and passive beamforming \cite{TCOM_2020_Wu_Beamforming} in Table \ref{tab1}. Note that an SIM eliminates the need for digital beamforming and operates directly on the EM waves. In addition, the SIM analyzed in this paper is different from a conventional reflecting RIS in several key aspects: (i) an SIM operates by diffracting rather than reflecting signals; (ii) an SIM is an integral component of the BS, hence resulting in a better path-loss scaling law \cite{WC_2022_An_Codebook}; and (iii) an SIM modifies both the amplitude and phase of the EM waves propagating through it, thus achieving better performance for mitigating interference.

To elaborate, the main contributions of this paper are summarized as follows:
\begin{enumerate}
\item An SIM-based transceiver is designed for the downlink of a multiuser MISO wireless system. In particular, an SIM is deployed to enhance the downlink communication from a multiple-antenna BS to multiple single-antenna user equipments (UEs). Thanks to the remarkable signal processing capabilities offered by the multilayer metasurface structure, the SIM allows for precise transmit beamforming in the EM wave domain. As a result, conventional beamforming schemes based on digital signal processing and high-resolution DACs are completely eliminated from the BS.
\item We formulate an optimization problem aiming at maximizing the sum rate of all UEs via jointly optimizing the transmit power allocation at the BS and the wave-based beamforming at the SIM, subject to a total transmit power budget and discrete phase shifts for the transmission coefficients. The formulated joint optimization problem, however, turns out to be a mixed integer non-linear programming (MINLP) problem, where the transmit power allocation coefficients and the discrete phase shifts are deeply coupled within the non-convex objective function, rendering it challenging to obtain the optimal solution.
\item To tackle this challenge, a computationally efficient alternating optimization (AO) algorithm is proposed to decompose the joint power allocation and wave-based beamforming optimization problem into two nested subproblems. The first subproblem deals with the conventional power allocation for multiuser interference channels, which can be effectively solved by modifying the iterative water-filling algorithm. The second subproblem involves optimizing the phase shifts of the SIM. To address this, we propose a pair of suboptimal algorithms including the projected gradient ascent algorithm and the successive refinement method.
\item We further provide useful insights into the potential benefits of employing SIM-based transceivers. Finally, numerical results are provided to evaluate the efficacy of the wave-based beamforming scheme, taking into account discrete phase shifts. Moreover, the enhanced performance offered by the proposed AO algorithm in comparison to various benchmark schemes is substantiated.
\end{enumerate}

The remainder of this paper is structured as follows. Section \ref{sec2} introduces the system model. Then, Section \ref{sec3} addresses the joint power allocation and wave-based beamforming problem. Furthermore, Section \ref{sec6} provides numerical results to evaluate the performance of the proposed wave-based beamforming scheme. Finally, Section \ref{sec7} concludes the paper.
\begin{figure*}[!t]
\centering
\includegraphics[width=16cm]{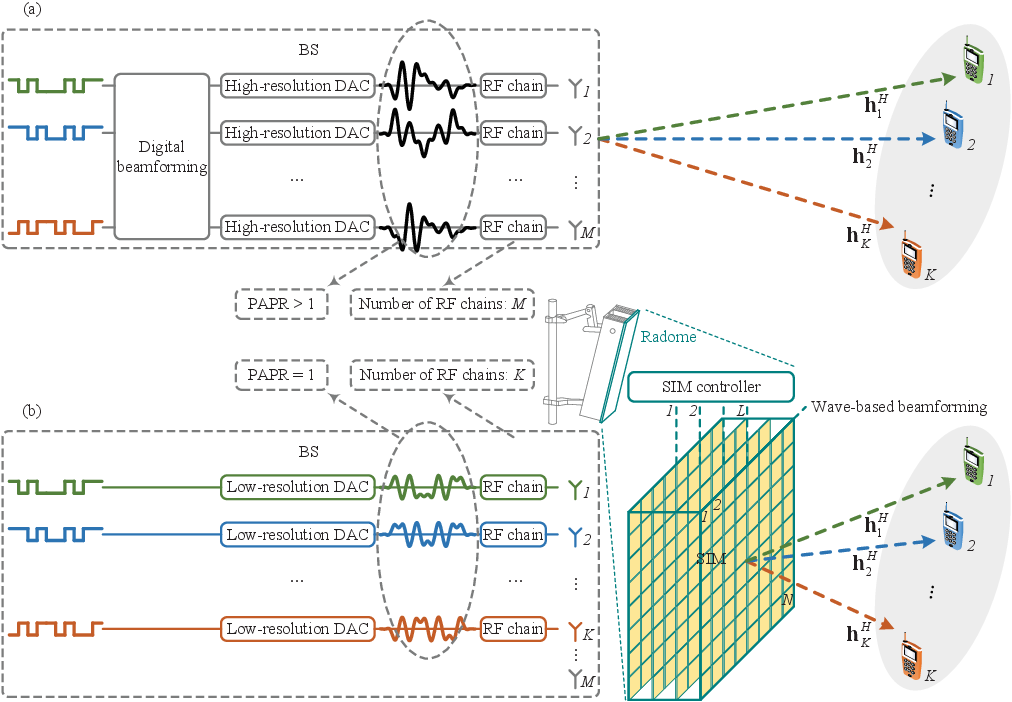}
\caption{Comparison of (a) conventional and (b) SIM-based multiuser MISO wireless systems, where binary phase shift keying modulation is adopted.}\vspace{-0.5cm}
\label{fig_1}
\end{figure*}

\emph{Notations:} Scalars are denoted by italic letters; column vectors and matrices are represented by boldface lower-case and upper-case symbols, respectively; $j$ is the imaginary unit satisfying $j^2=-1$. Moreover, $\left | z \right |$, $\Re \left ( z \right )$, and $\Im \left ( z \right )$ denote the absolute value, real part, and imaginary part, respectively, of a complex number $z$; $\left ( \cdot \right )^{\text{T}}$ and $\left ( \cdot \right )^{\text{H}}$ denote the transpose and Hermitian transpose, respectively; $\text{diag}\left ( x_{1},x_{2},\cdots ,x_{N} \right )$ generates a diagonal matrix with $x_{1},x_{2},\cdots ,x_{N}$ being its diagonal entries; $\mathbf{0}$ denotes the all-zero vector of appropriate dimension, while $\mathbf{I}_{N}$ represents the $N\times N$ identity matrix; $\mathbb{C}^{x \times y}$ represents the space of $x \times y$ complex-valued matrices. Furthermore, $\left \lfloor \cdot \right \rfloor$, $\left \lfloor \cdot \right \rceil$, $\left \lceil \cdot \right \rceil$, and $\text{mod}\left ( \cdot \right )$ represent the floor, rounding, ceiling, and modulo operator, respectively, while ${{\log}_{a}}\left ( \cdot \right )$ is the logarithmic function with base $a$; $\frac{\partial f}{\partial x}$ denotes the partial derivative of a function $f$ with respect to $x$. The distribution of a circularly symmetric complex Gaussian random vector with a mean vector $\mathbf{v}$ and a covariance matrix $\mathbf{\Sigma }$ is expressed as $\sim \mathcal{CN}\left ( \mathbf{v},\mathbf{\Sigma } \right )$, where $\sim $ stands for “distributed as”; $\text{sinc}\left ( x \right ) = \frac{\sin\left ( \pi x \right )}{\pi x}$ is the normalized sinc function.
\section{System Model}\label{sec2}
As shown in Fig. \ref{fig_1}, we consider the downlink of a multiuser MISO wireless system, where an SIM is integrated with the radome of the BS to assist the communication from $M$ antennas to $K$ single-antenna UEs within a given frequency band. In particular, the SIM is fabricated by utilizing an array of $L$ metasurfaces, each consisting of $N$ meta-atoms\footnote{An identical number of meta-atoms on each metasurface is assumed for the sake of simplicity but not loss of generality.}. Moreover, the SIM is connected to a smart controller, such as a customized FPGA, which is capable of independently adjusting the phase shift of the EM waves transmitted through each meta-atom \cite{TCOM_2020_Wu_Beamforming, TCOM_20202_An_Low}. Accordingly, the forward propagation process in SIM resembles a fully connected ANN. By properly training the architecture, the SIM can directly perform the downlink beamforming in the wave domain \cite{NE_2022_Liu_A}. Let $\mathcal{L} = \left \{ 1, 2, \cdots , L \right \}$, $\mathcal{N} = \left \{ 1, 2, \cdots , N \right \}$, and $\mathcal{K} = \left \{ 1, 2, \cdots , K \right \}$ denote the sets of metasurfaces, meta-atoms on each metasurface layer, and UEs, respectively.

Specifically, the complex-valued gain of the $n$-th meta-atom on the $l$-th metasurface layer is represented by $\alpha _{n}^{l}e^{j\theta _{n}^{l}},\, \forall n \in \mathcal{N},\, \forall l \in \mathcal{L}$, where $\alpha _{n}^{l}\in \left [ 0,1 \right ]$ and $\theta _{n}^{l} \in \left [ 0,2\pi \right )$ are the corresponding amplitude and phase shift, respectively. In order to achieve higher energy efficiency, each meta-atom is generally designed to maximize the signal transmission \cite{TWC_2019_Huang_Reconfigurable, TGCN_2022_An_Joint}. As such, we assume $\alpha _{n}^{l} = 1,\, \forall n \in \mathcal{N},\, \forall l \in \mathcal{L}$ throughout this paper. Additionally, in practical implementations, the phase shift of each meta-atom is limited to a finite number of discrete values. We assume that the discrete phase shift set $\mathcal{B}$ is obtained by uniformly quantizing the interval $\left [ 0, 2\pi \right )$ into $2^{b}$ levels, with $b$ representing the number of control bits. Thus, the set of legitimate phase shift values for each meta-atom is given by
\begin{align}
 \mathcal{B} = \left \{ 0, \Delta_{\theta} , 2\Delta_{\theta} , \cdots , \left ( 2^{b} - 1 \right )\Delta_{\theta} \right \},
\end{align}
where $\Delta_{\theta} = 2\pi / 2^{b}$ represents the phase shift resolution. As a result, the diagonal phase shift matrix $\mathbf{\Phi}^{l}$ of the $l$-th metasurface layer can be written as
\begin{align}
 \mathbf{\Phi}^{l} = \text{diag}\left ( e^{j\theta _{1}^{l}}, e^{j\theta _{2}^{l}}, \cdots , e^{j\theta _{N}^{l}} \right ) \in \mathbb{C}^{N\times N},\quad \forall l\in \mathcal{L},
\end{align}
where $\theta _{n}^{l} \in \mathcal{B},\, \forall n\in \mathcal{N},\, \forall l\in \mathcal{L}$.

Furthermore, let $\mathbf{W}^{l}\in \mathbb{C}^{N\times N},\, \forall l \neq 1,\, l \in \mathcal{L}$ represent the transmission matrix from the $\left ( l-1 \right )$-th to the $l$-th metasurface layer, and let $\mathbf{w}_{m}^{1}\in \mathbb{C}^{N\times 1}$ denote the transmission vector from the $m$-th transmit antenna to the first metasurface layer of the SIM. According to Rayleigh-Sommerfeld's diffraction theory \cite{NE_2022_Liu_A, Science_2018_Lin_All, BOOK_2005_Goodman_Introduction}, the $\left ( n,{n}' \right )$-th entry $w_{n,{n}'}^{l}$ of $\mathbf{W}^{l}$ is given by
\begin{align}\label{eq3}
 w_{n,{n}'}^{l} =\frac{d_{\text{x}}d_{\text{y}}\cos\chi _{n,{n}'}^{l} }{d_{n,{n}'}^{l}}\left ( \frac{1}{2\pi d_{n,{n}'}^{l}}-j\frac{1}{\lambda } \right )e^{j2\pi d_{n,{n}'}^{l}/\lambda },
\end{align}
for $\forall l \in \mathcal{L}$, where $\lambda$ is the wavelength, $d_{n,{n}'}^{l}$ denotes the transmission distance of the corresponding link, $\chi _{n,{n}'}^{l}$ represents the angle between the propagation direction and the normal direction of the $\left ( l-1 \right )$-th metasurface layer, while $d_{\text{x}} \times d_{\text{y}}$ characterizes the size of each meta-atom. Again, as the SIM is integrated within the transmitting antenna array, the $n$-th entry $w_{n,m}^{1}$ of $\mathbf{w}_{m}^{1}$ can also be obtained using \eqref{eq3} upon replacing $d_{n,{n}'}^{l}$ and $\chi _{n,{n}'}^{l}$ with $d_{n,m}^{1}$ and $\chi _{n,m}^{1}$, respectively.

According to the Huygens–Fresnel principle \cite{Science_2018_Lin_All}, the EM wave passing through each meta-atom on a given metasurface layer acts as a point source and illuminates all the meta-atoms on the subsequent metasurface layer. Furthermore, all EM waves impinging on a meta-atom of a metasurface layer are superimposed and then act as an incident wave onto the considered meta-atom. As such, the forward propagation in the SIM can be expressed as
\begin{align}\label{eq4}
 \mathbf{G}=\mathbf{\Phi }^{L}\mathbf{W}^{L}\mathbf{\Phi }^{L-1}\cdots \mathbf{\Phi }^{2}\mathbf{W}^{2}\mathbf{\Phi }^{1}\in \mathbb{C}^{N\times N}.
\end{align}
In practical implementations, an SIM is typically housed within a metallic support structure that is coated with wave-absorbing material \cite{NE_2022_Liu_A}. Hence, any diffraction effects between non-adjacent metasurfaces can be considered negligible and are disregarded in \eqref{eq4}.

In this paper, we consider a quasi-static flat-fading model for all channels illustrated in Fig. \ref{fig_1} (b). Specifically, let $\mathbf{h}_{k}^{\text{H}} \in \mathbb{C}^{1\times N},\, \forall k \in \mathcal{K}$ denote the baseband equivalent channel from the last metasurface layer to the $k$-th UE, which is modeled according to a correlated Rayleigh fading distribution \cite{CL_2023_An_A}, as follows: 
\begin{align}\label{eq5}
 \mathbf{h}_{k}\sim \mathcal{CN}\left ( \mathbf{0}, \beta _{k}\mathbf{R} \right ),\quad \forall k\in \mathcal{K},
\end{align}
where $\beta _{k}$ represents the distance-dependent path loss of the link between the SIM and UE $k$, while $\mathbf{R}\in \mathbb{C}^{N\times N}$ is a positive semidefinite matrix that characterizes the spatial correlation among the channels associated with different meta-atoms on the last layer. Assuming an isotropic scattering environment with uniformly distributed multipath components, the $\left ( n,{n}' \right )$-th entry of $\mathbf{R}$ can be expressed as \cite{WCL_2021_Bjornson_Rayleigh}
\begin{align}\label{eq6}
 \mathbf{R}_{n,{n}'}=\text{sinc} \left ( \frac{2d_{n,{n}'}}{\lambda} \right ),
\end{align}
where $d_{n,{n}'}$ represents the corresponding meta-atom spacing.

Next, we consider the downlink data transmission by adopting the spatial division multiple access (SDMA). As shown in Fig. \ref{fig_1} (a), the conventional SDMA relies on digital precoding that assigns each symbol an individual beamforming vector. Hence, each antenna transmits the superimposed signal of multiple users, which results in a high peak-to-average power ratio (PAPR). In contrast, Fig. \ref{fig_1} (b) employs wave-based beamforming with the aid of an SIM, where the matrix multiplication naturally occurs as the transmitted signals propagate through the SIM. As a result, each data stream is transmitted directly from the corresponding antenna at the BS, which implies that the BS needs to select a set of $K$ appropriate antennas from the total of $M$ antennas in advance. This antenna selection process resembles that in conventional MIMO systems with a limited number of RF chains at the BS \cite{CM_2004_Sanayei_Antenna, MM_2004_Molisch_MIMO}. Since this paper focuses on wave-based beamforming, we assume that $M = K$ for the sake of simplicity. The joint antenna selection and wave-based beamforming design are postponed to a future study.

Specifically, let $s_k,\, \forall k\in \mathcal{K}$ denote the information-bearing symbol to the $k$-th UE, which is modeled by an independent and identically distributed (i.i.d.) sequence of random variables with zero means and unit variances. Let $p_{k} \geq 0$ represent the power allocated to the $k$-th UE. Accordingly, the total transmit power constraint at the BS can be expressed as
\begin{align}
 \sum_{k = 1}^{K}p_{k} \leq P_{\text{T}},
\end{align}
where $P_{\text{T}}$ represents the transmit power budget at the BS. Furthermore, upon superimposing all the signals propagating through the SIM, the composite signal $r_{k}$ received at the $k$-th UE is given by
\begin{align}
 r_{k} = \mathbf{h}_{k}^{\text{H}}\mathbf{G}\sum\limits_{{k}' = 1}^{K}\mathbf{w}_{{k}'}^{1}\sqrt{p_{{k}'}}s_{{k}'} + n_{k},\quad \forall k \in \mathcal{K},
\end{align}
where $n_{k}\sim \mathcal{CN}\left ( 0,\sigma _{k}^{2} \right )$ denotes the additive white Gaussian noise (AWGN) with $\sigma _{k}^{2}$ representing the average noise power at the receiver of UE $k$.

By treating the signals of the remaining $\left ( K-1 \right )$ UEs as interference, the signal-to-interference-plus-noise-ratio (SINR) at the receiver of UE $k$ can be expressed as
\begin{align}\label{eq9}
 \gamma _{k} = \frac{\left | \mathbf{h}_{k}^{\text{H}}\mathbf{G}\mathbf{w}_{k}^{1} \right |^{2}p _{k}}{\sum\limits_{{k}'\neq k}^{K}\left | \mathbf{h}_{k}^{\text{H}}\mathbf{G}\mathbf{w}_{{k}'}^{1} \right |^{2}p _{{k}'} + \sigma _{k}^{2}},\quad \forall k \in \mathcal{K}.
\end{align}
As a result, the sum rate of these $K$ UEs is given by
\begin{align}\label{eq10}
 R=\sum \limits_{k = 1}^{K}\log_{2}\left ( 1+\gamma _{k} \right ).
\end{align}

\begin{remark}
In \eqref{eq9}, although UE $k$ naively treats the signals of the other UEs as interference while demodulating $s_{k}$, it is possible to optimize the phase shifts in $\mathbf{G}$ to mitigate the inter-user interference, which constitutes the proposed \emph{``wave-based beamforming''} design. While the wave-based beamforming scheme may appear to require a large number of matrix multiplications at the first glance, it is important to highlight that all the matrix multiplications in \eqref{eq4} are automatically executed as the wave propagates through each metasurface layer. This new wave-based computing paradigm is fundamentally distinct from baseband digital beamforming.
\end{remark}

\begin{remark}
In the considered system, the precoding vectors, i.e., the configuration of the meta-atoms of the SIM, are computed for every coherent block based on the available channel state information (CSI). When transmitting data, symbol-level precoding is not performed in the baseband through matrix multiplications, but directly in the wave domain as the signals propagate through the SIM.
\end{remark}

\begin{remark}
By employing the time-division duplex (TDD) protocol, the BS can readily acquire the downlink CSI by leveraging channel reciprocity. Since the number of RF chains is smaller than the number of meta-atoms, all the channels associated with $K$ UEs can be estimated by using at least $ \left \lceil KN/M \right \rceil$ pilot symbols \cite{WCL_2024_Yao_Channel}. Moreover, the phase shifts of the SIM can be jointly designed with the uplink pilots to further enhance the channel estimation accuracy \cite{arXiv_2023_Nadeem_Hybrid}.
\end{remark}

\begin{remark}
Due to practical hardware imperfections, such as the bending of metasurfaces, the transmission coefficients $w_{n,{n}'}^{l}$ between adjacent metasurface layers may deviate from the values specified in \eqref{eq3}. At the time of writing, there is no experimentally validated model available, which also leads to inevitable modeling errors in \eqref{eq3}. Therefore, it is necessary to calibrate the transmission coefficients $w_{n,{n}'}^{l}$ before practical SIM deployment. This calibration process can be accomplished by measuring the signals received at some auxiliary probes and applying the well-known backpropagation algorithm \cite{NE_2022_Liu_A}.
\end{remark}

\begin{remark}
The propagation loss through the SIM includes the penetration loss through each individual layer and the propagation attenuation between adjacent metasurface layers. For simplicity, the former has been ignored in this work, while the latter has been implicitly considered using the cascaded channel model in \eqref{eq4}. Additionally, the wave-based beamforming and the large aperture of the metasurfaces provide the BS addition power margin to compensate for these penetration and propagation losses. In general, however, a holistic system-level performance evaluation of SIM-aided communication systems compared to conventional transceivers requires further investigation.
\end{remark}

\begin{remark}
The mutual coupling between metasurface layers and meta-atoms would impact the EM response of the SIM \cite{NE_2022_Liu_A}. Accounting for the mutual coupling in the SIM design may further enhance the performance of the SIM, but this requires accurate modeling. In this paper, we assume a negligible coupling effect by placing the layers and meta-atoms at least half a wavelength apart.
\end{remark}

\section{Joint Power Allocation and Wave-Based Beamforming: Problem Formulation and Solution}\label{sec3}
\subsection{Problem Formulation}\label{sec3-1}
We aim to maximize the sum rate of all UEs by jointly optimizing the transmit power allocation at the BS and the wave-based beamforming at the SIM, subject to constraints on the total transmit power budget and the discrete phase shifts. To characterize the ultimate performance limits, we assume that the CSI of all the channels in Fig. \ref{fig_1} (b) is perfectly known to the BS. By introducing the variables $\mathbf{p} \triangleq \left [ p_{1},p_{2},\cdots ,p_{K} \right ]^{\text{T}}\in \mathbb{C}^{K\times 1}$, ${\bm{\theta }}^{l} \triangleq \left [ \theta _{1}^{l}, \theta _{2}^{l}, \cdots , \theta _{N}^{l} \right ]^{\text{T}}\in \mathbb{C}^{N\times 1}$, and $\boldsymbol{\vartheta } \triangleq \left \{ \boldsymbol{\theta} ^{1},\boldsymbol{\theta} ^{2},\cdots ,\boldsymbol{\theta} ^{L} \right \}$, the joint power allocation and wave-based beamforming optimization problem is formulated as
\begin{subequations}
\begin{alignat}{3}
\left ( \mathcal{P_{A}} \right ): \ & \max_{\mathbf{p},\, \boldsymbol{\vartheta }} \ && R=\sum \nolimits_{k = 1}^{K}\log_{2}\left ( 1+\gamma _{k} \right ) \label{eq11a}\\
{} \ & \text{s.t.} \ && \gamma _{k} = \frac{\left | \mathbf{h}_{k}^{\text{H}}\mathbf{G}\mathbf{w}_{k}^{1} \right |^{2}p _{k}}{\sum\nolimits_{{k}'\neq k}^{K}\left | \mathbf{h}_{k}^{\text{H}}\mathbf{G}\mathbf{w}_{{k}'}^{1} \right |^{2}p _{{k}'} + \sigma _{k}^{2}},\ \forall k, \label{eq11b}\\
{} \ & {} \ && \mathbf{G}=\mathbf{\Phi }^{L}\mathbf{W}^{L}\mathbf{\Phi }^{L-1}\cdots \mathbf{\Phi }^{2}\mathbf{W}^{2}\mathbf{\Phi }^{1}, \label{eq11c}\\
{} \ & {} \ && \mathbf{\Phi}^{l} = \text{diag}\left ( e^{j\theta _{1}^{l}}, e^{j\theta _{2}^{l}}, \cdots , e^{j\theta _{N}^{l}} \right ),\ \forall l, \label{eq11d}\\
{} \ & {} \ && \theta _{n}^{l} \in \mathcal{B},\ \forall n \in \mathcal{N},\ \forall l \in \mathcal{L}, \label{eq11e}\\
{} \ & {} \ && \sum\nolimits_{k=1}^{K}p_{k}\leq P_{\text{T}}, \label{eq11f}\\
{} \ & {} \ && p_{k}\geq 0,\ \forall k \in \mathcal{K}. \label{eq11g}
\end{alignat}
\end{subequations}
Despite the conciseness of problem $\left ( \mathcal{P_{A}} \right )$, it is generally challenging to obtain an optimal solution for the joint power allocation and wave-based beamforming design, due to the fact that the optimization variables $\mathbf{p}$ and $\boldsymbol{\vartheta }$ are deeply coupled within the non-convex objective function \eqref{eq11a}. In addition, the constraints in \eqref{eq11e} restrict the phase shifts $\theta _{n}^{l}$ to take discrete values, thereby rendering problem $\left ( \mathcal{P_{A}} \right )$ an NP-hard MINLP. To address this challenge, we propose an AO algorithm in Section \ref{sec3-2} to obtain a suboptimal solution for problem $\left ( \mathcal{P_{A}} \right )$.

\begin{remark}
SIM-aided MIMO systems have been examined in \cite{JSAC_2023_An_Stacked} by formulating a channel fitting problem to minimize the error between the SIM-aided end-to-end channel and the desired diagonal channel based on singular value decomposition. By contrast, in this paper, we aim to optimize the SIM phase shifts to maximize the system sum rate while taking into account practical hardware constraints.
\end{remark}
\subsection{The Proposed Alternating Optimization Algorithm}\label{sec3-2}
The AO algorithm decomposes the joint optimization problem $\left ( \mathcal{P_{A}} \right )$ into a pair of nested subproblems: one involving the conventional power allocation among $K$ interference channels, and the other focusing on the phase shift optimization. In each subproblem, only one of the optimization variables is updated, either $\mathbf{p}$ or $\boldsymbol{\vartheta }$, while keeping the other one fixed at the value obtained at the previous iteration.

\subsubsection{Optimization of the Power Allocation $\mathbf{p}$ Given $\boldsymbol{\vartheta }$}\label{sec41}
For any given SIM phase shifts $\boldsymbol{\vartheta } = \left \{ \boldsymbol{\theta} ^{1},\boldsymbol{\theta} ^{2},\cdots ,\boldsymbol{\theta} ^{L} \right \}$, the wave-based beamforming matrix $\mathbf{G}$ can be determined by utilizing \eqref{eq4}. As such, problem $\left ( \mathcal{P_{A}} \right )$ is simplified to
\begin{subequations}
\begin{alignat}{3}
\left ( \mathcal{P_{B}} \right ): \quad & \max_{\mathbf{p}} \quad && R\\
{} \quad & \text{s.t.} \quad && \eqref{eq11b},\ \eqref{eq11f},\ \eqref{eq11g}.
\end{alignat}
\end{subequations}
Problem $\left ( \mathcal{P_{B}} \right )$ turns out to be the traditional power allocation problem among $K$ interference channels with a given total power budget, which can be efficiently solved by employing the iterative water-filling algorithm \cite{TIT_2004_Wei_Iterative, TIT_2005_Jindal_Sum}.

To elaborate, once an initial power allocation (e.g., average power allocation) is chosen, the interference power perceived by each UE can be calculated accordingly. At each iteration, UE $k$ treats the interference caused by all the other UEs as background noise. Accordingly, the optimal power allocation solution for maximizing the sum rate can be obtained by employing the well-known water-filling principle as \cite{BOOK_2005_Tse_Fundamentals}
\begin{align}\label{eq13}
 p_{k}=\left ( p_{\text{WF}}-\frac{\sum\nolimits_{{k}'\neq k}^{K}\left | \mathbf{h}_{k}^{\text{H}}\mathbf{G}\mathbf{w}_{{k}'}^{1} \right |^{2}p _{{k}'} + \sigma _{k}^{2}}{\left | \mathbf{h}_{k}^{\text{H}}\mathbf{G}\mathbf{w}_{k}^{1} \right |^{2}} \right )^{+},
\end{align}
for $\forall k\in \mathcal{K}$, where $\left ( x \right )^{+} \triangleq \max\left \{ 0,x \right \}$, while the water-filling level $p_{\text{WF}}$ is determined using the bisection method such that $\sum\nolimits_{k=1}^{K}p_{k}=P_{\text{T}}$ is satisfied at each iteration.

Nevertheless, as the power allocated to all the UEs is simultaneously updated at each iteration, the typical iterative water-filling algorithm in \eqref{eq13} may not be stable when $K > 2$ \cite{TIT_2004_Wei_Iterative}. To address this issue, we modify the update process by introducing a damping term. Specifically, the updated power allocation solution $\mathbf{p}$ at each iteration is a weighted combination of the previous power allocation solution and the new one that is generated by the iterative water-filling algorithm. Let $\mathbf{p}^{\left ( \star \right )}$ represent the new power allocation solution obtained using \eqref{eq13}. Hence, we have
\begin{align}\label{eq14}
 \mathbf{p}^{\left ( i \right )}=\zeta \mathbf{p}^{\left ( \star \right )}+\left ( 1-\zeta \right )\mathbf{p}^{\left ( i-1 \right )},
\end{align}
where $\mathbf{p}^{\left ( i-1 \right )}$ and $\mathbf{p}^{\left ( i \right )}$ represent the power allocation solutions at the $\left ( i-1 \right )$-th and the $i$-th iteration, respectively, while $1/K\leq \zeta\leq 1$ is a weighting coefficient controlling the update step size.

By repeatedly applying \eqref{eq13} and \eqref{eq14} multiple times, the sum rate $R$ will gradually approach its maximum value. In \emph{Theorem 3} of \cite{TIT_2005_Jindal_Sum}, Jindal \emph{et al.} have proven that the modified iterative water-filling algorithm incorporating a damping term is guaranteed to converge to the sum rate capacity for arbitrary values of $K$. The convergence of the modified iterative water-filling algorithm will be numerically illustrated in Section \ref{sec6}.

\subsubsection{Optimization of the SIM Phase Shifts\label{sec42} $\boldsymbol{\vartheta }$ Given $\mathbf{p}$}
Given a tentative power allocation solution $\mathbf{p}$, the phase shift optimization subproblem is formulated as
\begin{subequations}
\begin{alignat}{3}
\left ( \mathcal{P_{C}} \right ): \quad & \max_{\boldsymbol{\vartheta }} \quad && R \label{eq15a}\\
{} \quad & \text{s.t.} \quad && \eqref{eq11b} - \eqref{eq11e}. \label{eq15b}
\end{alignat}
\end{subequations}
In general, the globally optimal phase shifts for problem $\left ( \mathcal{P_{C}} \right )$ can only be obtained via exhaustively searching through all the legitimate phase shift values. As a result, the total computational complexity is given by $\mathcal{O}\left [ \left ( 4N+3 \right )K^{2}B^{LN} \right ]$, where $\left ( 4N+3 \right )$ represents the number of real-valued multiplications required to compute $\left | \mathbf{h}_{k}^{\text{H}}\mathbf{G}\mathbf{w}_{{k}'}^{1} \right |^{2}p _{{k}'}$ in \eqref{eq11b} for each pair $\left ( k,k' \right )\in \mathcal{K}^{2}$, bearing in mind that $\mathbf{G}\mathbf{w}_{{k}}^{1} \in \mathbb{C}^{N\times 1}$ is known \emph{a priori} for each $\theta _{n}^{l}\in \mathcal{B}$. Note that the computational complexity of the exhaustive search grows exponentially with the number of metasurface layers $L$ and with the number of meta-atoms $N$ on each layer, which is prohibitive even for a medium-sized SIM. Next, we propose a pair of computationally efficient algorithms, namely the projected gradient ascent and the successive refinement methods, to solve problem $\left ( \mathcal{P_{C}} \right )$.

\emph{i) Successive Refinement Method:}
The suboptimal successive refinement algorithm iteratively optimizes each, out of the $LN$ phase shifts to be optimized, once at a time. Specifically, by keeping fixed $\left ( LN-1 \right )$ phase shifts, the optimal solution for the $n$-th discrete phase shift $\theta _{n}^{l}$ on the $l$-th metasurface layer can be obtained via a one-dimensional search over $\mathcal{B}$, i.e.,
\begin{align}\label{eq17}
\hat{\theta }_{n}^{l}=\arg\underset{\theta _{n}^{l}\in \mathcal{B}}{\max}\ R,\quad \forall n \in \mathcal{N},\quad \forall l \in \mathcal{L}.
\end{align}

The proposed successive refinement algorithm is guaranteed to converge due to the following two reasons. Firstly, by successively updating the phase shifts of all meta-atoms based on \eqref{eq17}, the objective function value of $\left ( \mathcal{P_{C}} \right )$ in \eqref{eq15a} is non-decreasing during the iterative process \cite{JSAC_2020_Di_Hybrid}. Secondly, the optimal objective value of $\left ( \mathcal{P_{C}} \right )$ is upper-bounded as
\begin{align}\label{eq22}
 R&\overset{\left ( a \right )}{\leq} \sum \limits_{k = 1}^{K}\log_{2}\left ( 1+\underset{\boldsymbol{\vartheta} }{\max}\left | \mathbf{h}_{k}^{\text{H}}\mathbf{G}\mathbf{w}_{k}^{1} \right |^{2}\frac{p _{k}}{\sigma _{k}^{2}} \right ),
\end{align}
where $\left ( a \right )$ holds due to the finite number of discrete phase shift values with unit (finite) amplitude. Using the discrete phase shifts obtained at convergence, the maximum sum rate $R$ can be attained.

\emph{ii) Projected Gradient Ascent Method:}
Given that the optimization problem $\left ( \mathcal{P_{C}} \right )$ is a constrained maximization problem, the projected gradient ascent algorithm can be employed to iteratively update the phase shifts $\boldsymbol{\vartheta }$ until converging to a stationary point. The specific steps of the projected gradient ascent algorithm are outlined as follows:

\textbf{\emph{Step 1:}} Initialize all phase shifts $\theta _{n}^{l},\, \forall n\in \mathcal{N},\, \forall l \in \mathcal{L}$ to feasible values according to the feasible set $\mathcal{B}$. Then, calculate the sum rate $R$ by applying \eqref{eq10}.

\textbf{\emph{Step 2:}} Compute the partial derivatives of $R$ with respect to $\theta _{n}^{l}$ according to \emph{Proposition 1}.

\emph{Proposition 1:} For $\forall n\in \mathcal{N},\, \forall l \in \mathcal{L}$, the partial derivative of $R$ with respect to $\theta _{n}^{l}$ is as follows:
\begin{align}\label{eq15}
 \frac{\partial R}{\partial \theta _{n}^{l}}&=2\log_{2}e\sum_{k=1}^{K}\delta _{k} \left ( p_{k}\eta _{n,k,k}^{l}-\gamma _{k}\sum\limits_{{k}'\neq k}^{K}p_{{k}'}\eta _{n,k,{k}'}^{l} \right ),
\end{align}
where $\delta _{k}$ and $\eta _{n,k,{k}'}^{l}$ are defined by
\begin{align}
 \delta _{k}&\triangleq\frac{1}{\sum\nolimits_{{k}' = 1}^{K}\left | \mathbf{h}_{k}^{\text{H}}\mathbf{G}\mathbf{w}_{{k}'}^{1} \right |^{2}p _{{k}'} + \sigma _{k}^{2}},\label{eqq19}\\
 \eta _{n,k,{k}'}^{l}&\triangleq\Im \left [ e^{-j\theta _{n}^{l}}\left ( \mathbf{w}_{{k}'}^{1} \right )^{\text{H}}\mathbf{u}_{n}^{l}\left ( \mathbf{v}_{n}^{l} \right )^{\text{H}}\mathbf{h}_{k}\mathbf{h}_{k}^{\text{H}}\mathbf{G}\mathbf{w}_{{k}'}^{1} \right ],\label{eqq20}
\end{align}
respectively. Moreover, $\left ( \mathbf{u}_{n}^{l} \right )^{\text{H}}$ and $\mathbf{v}_{n}^{l}$ represent the $n$-th row of $\mathbf{U}^{l}\in \mathbb{C}^{N\times N}$ and the $n$-th column of $\mathbf{V}^{l}\in \mathbb{C}^{N\times N}$, which are defined by
\begin{align}
\mathbf{U}^{l}&\triangleq \begin{cases}
 \mathbf{W}^{l}\mathbf{\Phi }^{l-1}\cdots \mathbf{\Phi }^{2}\mathbf{W}^{2}\mathbf{\Phi }^{1}, & \text{if}\ l \neq 1,\\
 \mathbf{I}_{N}, & \text{if}\ l=1,
 \end{cases}\label{eqq21}\\
\mathbf{V}^{l}&\triangleq\begin{cases}
 \mathbf{\Phi }^{L}\mathbf{W}^{L}\mathbf{\Phi }^{L-1}\cdots \mathbf{\Phi }^{l+1}\mathbf{W}^{l+1}, & \text{if}\ l\neq L,\\
 \mathbf{I}_{N}, & \text{if}\ l=L,
 \end{cases}\label{eqq22}
\end{align}
respectively.

\emph{Proof:} Please refer to Appendix \ref{A1}. $\hfill \blacksquare$

\textbf{\emph{Step 3:}} After calculating all the partial derivatives of $R$ with respect to $\theta _{n}^{l}$ based on \eqref{eq15}, the phase shifts $\theta _{n}^{l}$ are simultaneously updated as follows:
\begin{align}\label{eq23}
 \theta _{n}^{l}\leftarrow \theta _{n}^{l}+\mu \frac{\partial R}{\partial \theta _{n}^{l}},\quad \forall n \in \mathcal{N},\quad \forall l \in \mathcal{L},
\end{align}
where $\mu > 0$ is the \emph{Armijo} step size, which is determined by leveraging the backtracking line search at each iteration \cite{TWC_2022_Papazafeiropoulos_Intelligent}.

\textbf{\emph{Step 4:}} Since the phase shift vector computed with \eqref{eq23} may not be a feasible solution due to the constraint in \eqref{eq11e}, a projection is performed to quantize $\theta _{n}^{l}$ into its nearest value in $\mathcal{B}$, i.e.,
\begin{align}
 \theta _{n}^{l}&\leftarrow\arg\underset{\theta \in \mathcal{B}}{\min}\ \left | \theta -\theta _{n}^{l} \right | \notag\\ &=\Delta_{\theta} \left \lfloor \theta _{n}^{l}/\Delta_{\theta} \right \rceil,\quad \forall n \in \mathcal{N},\quad \forall l \in \mathcal{L},
\end{align}
bearing in mind that $\Delta_{\theta} = 2\pi / 2^{b}$.

\textbf{\emph{Step 5:}} Repeat \textbf{\emph{Steps 2 $\sim$ 4}} until the fractional increase of the sum rate becomes smaller than a preset threshold. Then, return the corresponding $\theta _{n}^{l},\, \forall n\in \mathcal{N},\, \forall l \in \mathcal{L}$ as the optimized phase shifts of the SIM.

\begin{algorithm}[!t]
\caption{Alternating Optimization Algorithm for Joint Power Allocation and Wave-Based Beamforming.}
\label{alg1}
\begin{algorithmic}[1]
\STATE {\textbf{Input:}} $\left \{ \mathbf{W}^{l} \right \}_{l=1}^{L}$, $\left \{ \mathbf{h}_{k}^{\text{H}} \right \}_{k=1}^{K}$, $P_{\text{T}}$, and $\mathcal{B}$.
\STATE Initialize $\boldsymbol{\vartheta }$ that fulfills \eqref{eq11e}.
\STATE {\textbf{Repeat}}
\STATE \hspace{0.5cm} Update $\mathbf{p}$ by employing the modified iterative water-\\
\hspace{0.5cm} filling algorithm.
\STATE \hspace{0.5cm} Update $\boldsymbol{\vartheta }$ by employing the projected gradient ascent\\
\hspace{0.5cm} or successive refinement method.
\STATE {\textbf{Until}} The sum rate $R$ in \eqref{eq10} converges.
\STATE {\textbf{Output:}} $\mathbf{p}$ and $\boldsymbol{\vartheta }$.
\end{algorithmic}
\end{algorithm}
The convergence of the projected gradient ascent algorithm to a local maximum is guaranteed due to the fact that: \emph{i)} the sum rate $R$ is upper bounded by \eqref{eq22} and \emph{ii)} the sum rate $R$ is non-decreasing by choosing an appropriate \emph{Armijo} step size $\mu$ at each iteration \cite{PJM_1966_Armijo_Minimization}. Nevertheless, since the quantization process at each iteration might impact the performance, multiple initializations are employed to enhance the reliability of the projected gradient ascent algorithm.

As a result, a suboptimal solution for $\left ( \mathcal{P_{A}} \right )$ can be obtained by iteratively carrying out the optimization procedures outlined in Section \ref{sec3-2}. More explicitly, the complete AO algorithm solving problem $\left ( \mathcal{P_{A}} \right )$ is summarized in Algorithm \ref{alg1}.

\subsection{Computational Complexity Analysis}\label{sec4_3}
The proposed AO algorithm is a nested iterative optimization process. Specifically, the outer loop tackles two subproblems for optimizing $\mathbf{p}$ and $\boldsymbol{\vartheta }$, respectively, and each subproblem requires its own iterative updating method to find a solution. As for the power allocation, in particular, the complexity of the iterative water-filling algorithm in Section \ref{sec3-2} is $\mathcal{O}\left [ K^{2}I_{\text{IWF}} \left ( 4N+3 \right ) \right ]$, where $I_{\text{IWF}}$ denotes the number of iterations required for achieving convergence. As for the phase shift optimization, the complexity of the successive refinement algorithm to determine the phase shifts of the SIM is $\mathcal{O}\left [ K^{2}LNBI_{\text{SR}} \left ( 4N+3 \right ) \right ]$, where $I_{\text{SR}}$ represents the number of iterations required by the successive refinement algorithm to converge. Since the set of discrete phase shifts $\mathcal{B}$ is typically small in practice, the successive refinement algorithm relying on the one-dimensional search in \eqref{eq17} is computationally efficient. Furthermore, the gradient ascent algorithm has a computational complexity of $\mathcal{O}\left [ 2K^{2}LNI_{\text{GA}}\left ( 4N+3 \right ) \right ]$, where the complexities of calculating \eqref{eq10}, \eqref{eq15}, and \eqref{eqq20} are $\mathcal{O}\left [ K^{2}\left ( 4N+3 \right ) \right ]$, $\mathcal{O}\left ( K^{2} \right )$, and $\mathcal{O}\left [ K^{2}\left ( 4N+2 \right ) \right ]$, respectively. Note that the complexity of the gradient ascent algorithm is independent of the cardinality of $\mathcal{B}$, making it more efficient when meta-atoms with high-resolution phase shifts are considered. As a result, the total complexity of the proposed AO algorithm can be expressed as $\mathcal{O}\left [ I_{\text{AO}} K^{2}\left ( I_{\text{IWF}}+LN\min\left ( BI_{\text{SR}},2I_{\text{GA}} \right ) \right ) \left ( 4N+3 \right ) \right ]$, where $I_{\text{AO}}$ denotes the number of iterations required by the outer loop.

\section{Numerical Results}\label{sec6}
In this section, numerical results are provided to validate the effectiveness of the wave-based beamforming design and evaluate the sum rate of an SIM-assisted multiuser downlink MISO system utilizing the proposed algorithms.
\begin{figure}[!t]
\centering
\includegraphics[width=8.8cm]{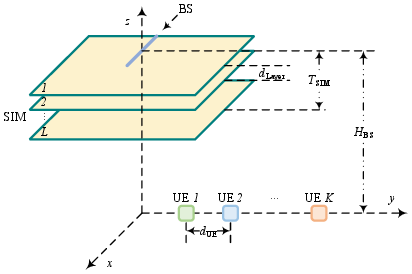}
\caption{Simulation setup for the considered SIM-assisted multiuser downlink MISO system.}\vspace{-0.5cm}
\label{fig_2}
\end{figure}
\subsection{Simulation Setup}
Fig. \ref{fig_2} illustrates the considered SIM-assisted downlink multiuser MISO system, where a BS is equipped with a uniform linear array (ULA) consisting of $M$ antennas positioned along the $\text{x}$-axis. Furthermore, an SIM is integrated with the BS to perform the transmit beamforming in the wave domain. Each metasurface of the SIM comprises a uniform planar array (UPA) that is deployed parallel to the $\text{x}$-$\text{y}$ plane. The center antenna/meta-atom of both the BS and metasurfaces are aligned with the $\text{z}$-axis. The height of the BS is set to $H_{\text{BS}} = 10$ m, and the SIM has a thickness of $T_{\text{SIM}} = 5\lambda$. Hence, the spacing between adjacent metasurfaces in an $L$-layer SIM is $d_{\text{Layer}} = T_{\text{SIM}}/L$. Moreover, we assume that all metasurfaces are isomorphic, with $N_{\text{x}}$ and $N_{\text{y}}$ representing the number of meta-atoms along the $\text{x}$-axis and the $\text{y}$-axis, respectively. Thus, we have $N=N_{\text{x}}N_{\text{y}}$. For the sake of simplicity, square metasurfaces are considered with $N_{\text{x}} = N_{\text{y}}$. Furthermore, a half-wavelength spacing between adjacent antennas/meta-atoms is assumed for both the BS and metasurfaces. The size of each meta-atom is $d_{\text{x}} = d_{\text{y}} = \lambda /2$. To mitigate the effects caused by the statistical positions of the UEs, $K$ single-antenna UEs are uniformly distributed along a line originating from the reference point with a spacing of $d_{\text{UE}} = 10$ m, as shown in Fig. \ref{fig_2}. Additionally, each antenna at the BS is assumed to have an antenna gain of $5$ dBi, while each UE is equipped with a single antenna with $0$ dBi gain \cite{TWC_2022_Papazafeiropoulos_Intelligent}.

Furthermore, the attenuation coefficients $w_{n,{n}'}^{l}$ between adjacent metasurface layers in the SIM and those from the transmit antenna array to the first layer of the SIM are computed by using \eqref{eq3}. Specifically, the distance between the ${n}'$-th meta-atom of the $\left ( l-1 \right )$-th metasurface and the $n$-th meta-atom of the $l$-th metasurface is given by $d_{n,{n}'}^{l} = \sqrt{d_{\text{Layer}}^{2}+d_{n,{n}'}^{2}}$, where $d_{n,{n}'}$ is defined as follows:
\begin{align}
d_{n,{n}'}=\frac{\lambda}{2}\sqrt{ \left \lfloor \left | n-{n}' \right |/N_{x} \right \rfloor^{2}+\left[\text{mod}\left ( \left | n-{n}' \right |,N_{x} \right ) \right ]^{2} }.
\end{align}
In particular, the transmission distance between the $m$-th antenna and the $n$-th meta-atom on the first layer of the SIM is determined by \eqref{eq26}\begin{figure*}
\begin{align}\label{eq26}
 d_{n,m}^{1} = \sqrt{d_{\text{Layer}}^{2}+ \left [ \left( \text{mod}\left ( n-1,N_{\text{x}} \right )-\frac{N_{\text{x}}-1 }{2} \right )-\left ( m-\frac{M+1 }{2} \right ) \right ]^{2}\frac{\lambda ^{2}}{4}+\left (\left \lceil n/N_{\text{x}} \right \rceil-\frac{ N_{\text{y}} +1 }{2} \right )^{2}\frac{\lambda ^{2}}{4}}.
\end{align}
\dotfill
\end{figure*}, as shown at the top of the next page. Furthermore, by definition, we have $\cos\chi _{n,{n}'}^{l} = d_{\text{Layer}}/d_{n,{n}'}^{l}$ for $\forall l\in \mathcal{L}$. The channels between the last layer of the SIM and the UEs are obtained from \eqref{eq5} with the spatial correlation matrix given in \eqref{eq6}. The distance-dependent path loss is modeled as
\begin{align}
 \beta_{k}=C_{0}\left ( d_{k}/d_{0} \right )^{-\bar{n}},\quad d_{k}\geq d_{0},
\end{align}
where $d_{k}=\sqrt{\left ( H_{\text{BS}}-T_{\text{SIM}} \right )^{2}+\left [ d _{\text{UE}}\left ( k-1 \right ) \right ]^{2}}$ represents the distance from the last layer of the SIM to the $k$-th UE, while $C_{0} = \left ( \lambda /4\pi d_{0} \right )^{2}$ is the free space path loss at the reference distance $d_{0} = 1$ m \cite{TCOM_2015_Rappaport_Wideband}, and the path loss exponent $\bar{n}$ is set to $\bar{n} = 3.5$. Additionally, we consider an indoor communication scenario operating at a carrier frequency of $28$ GHz with a transmission bandwidth of $10$ MHz and an effective noise power spectral density of $-174$ dBm/Hz for all UEs. Thus, we have $C_{0} = -40$ dB and $\sigma _{k}^{2} = - 104$ dBm for $\forall k \in \mathcal{K}$.

For comparison, we evaluate the performance of the proposed algorithms against the following four benchmark schemes:
\begin{itemize}
 \item[$\circ$] \emph{Continuous phase shift:} In this case, problem $\left ( \mathcal{P_{A}} \right )$ is solved by relaxing all discrete optimization variables $\theta_{n}^{l}$ to their continuous counterparts. To solve subproblem $\left ( \mathcal{P_{C}} \right )$, the projected gradient ascent method is adapted by excluding \textbf{\emph{Step 4}};
 \item[$\circ$] \emph{Quantization scheme:} In this case, each continuous phase shift is quantized to its nearest discrete value in $\mathcal{B}$;
 \item[$\circ$] \emph{Average power allocation:} In this case, problem $\left ( \mathcal{P_{C}} \right )$ is solved given the average power allocation solution;
 \item[$\circ$] \emph{Codebook method:} In this case, a random phase shift vector codebook of size $Q$ is generated. For each phase shift in the codebook, the iterative water-filling power allocation is applied. Subsequently, the phase shift that yields the maximum sum rate is selected.
\end{itemize}

Moreover, the condition for terminating the AO and the iterative water-filling algorithms is that the fractional increase of the sum rate is less than $10^{-6}$. Additionally, the maximum number of iterations for the AO algorithm, the inner iteration of the iterative water-filing, and the backtracking line search is $100$. All simulation results are obtained by averaging over $100$ channel realizations, each with independent small-scale fading. To avoid the exponentially growing solution space (i.e., $2^{bLN}$) with the total number of meta-atoms, a linear codebook size $Q = 10LN$ is considered. Each codeword of the codebook is obtained by randomly generating the phase shifts.

\subsection{Sum Rate Performance Evaluation of the Proposed Wave-Based Beamforming Design}
\begin{figure}[!t]
\centering
\includegraphics[width=8.5cm]{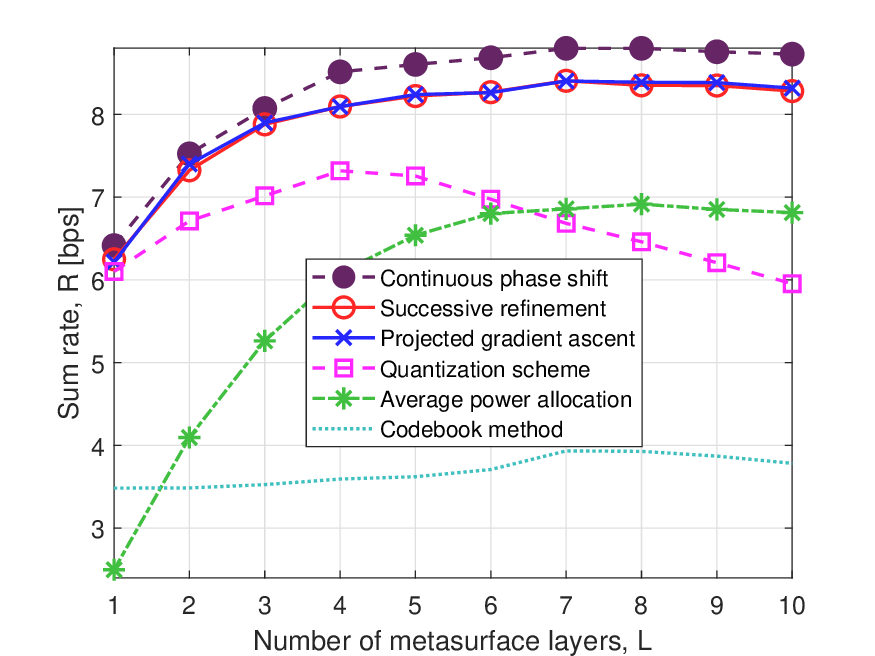}
\caption{Sum rate $R$ versus the number of metasurface layers $L$ ($N=49$, $b=2$, $M=K=4$, $P_{\text{T}}=10$ dBm).}
\label{fig_3}
\end{figure}

Fig. \ref{fig_3} illustrates the sum rate $R$ versus the number of metasurface layers $L$, when we set $N = 49$, $M = K = 4$, and $P_{\text{T}} = 10$ dBm. In particular, $b = 2$ is considered for the discrete phase shifts. By gradually increasing the number of metasurface layers from $L = 1$ to $L = 10$, the sum rate of the SIM-aided MISO system utilizing the optimized power allocation and wave-based beamforming improves, thanks to the ability of the SIM to mitigate the inter-user interference in the wave domain. In addition, the sum rate exhibits a tendency to converge and achieves its peak at $L = 7$, attaining a remarkable 130\% enhancement compared to a single-layer SIM. Assuming discrete phase shifts, the proposed SIM only experiences a minor performance loss of less than $0.5$ bps as compared to the case with continuous phase shifts. The performance gap can be further narrowed by increasing the value of $b$. Furthermore, the proposed algorithms outperform all benchmark schemes that consider discrete phase shifts. The quantization method initially exhibits a moderate performance erosion for a small number of metasurface layers, but it degrades significantly as $L > 4$ due to the quantization errors. The average power allocation solution suffers from a performance gap of $1.5$ bps in comparison to the iterative water-filling counterpart. Finally, the random phase shift fails to provide any noticeable performance gain as the number of metasurface layers increases.

\begin{figure}[!t]
\centering
\includegraphics[width=8.5cm]{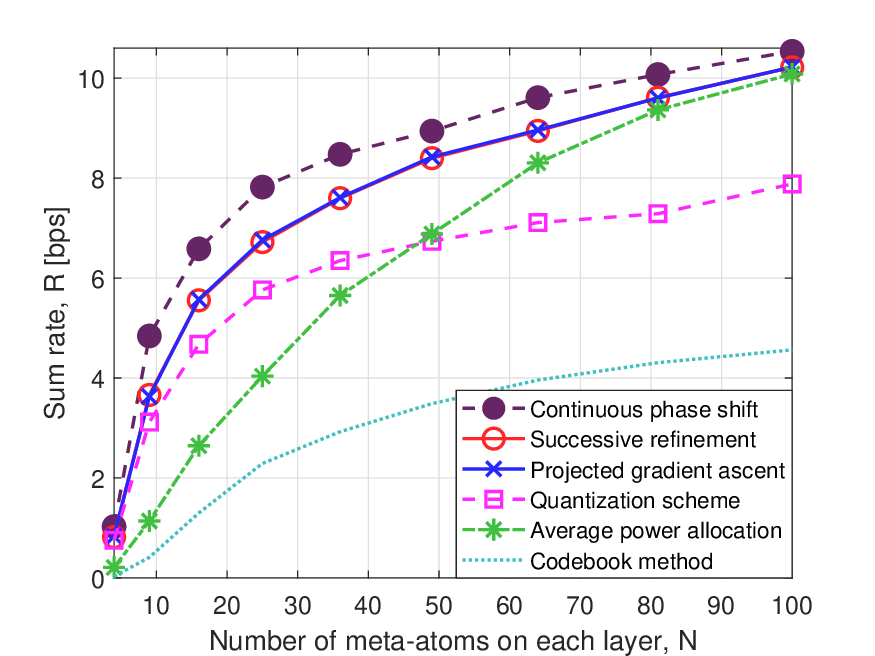}
\caption{Sum rate $R$ versus the number of meta-atoms $N$ on each metasurface layer ($b=2$, $L=7$, $M=K=4$, $P_{\text{T}}=10$ dBm).}\vspace{-0.5cm}
\label{fig_4}
\end{figure}

\begin{figure}[!t]
\centering
\includegraphics[width=8.5cm]{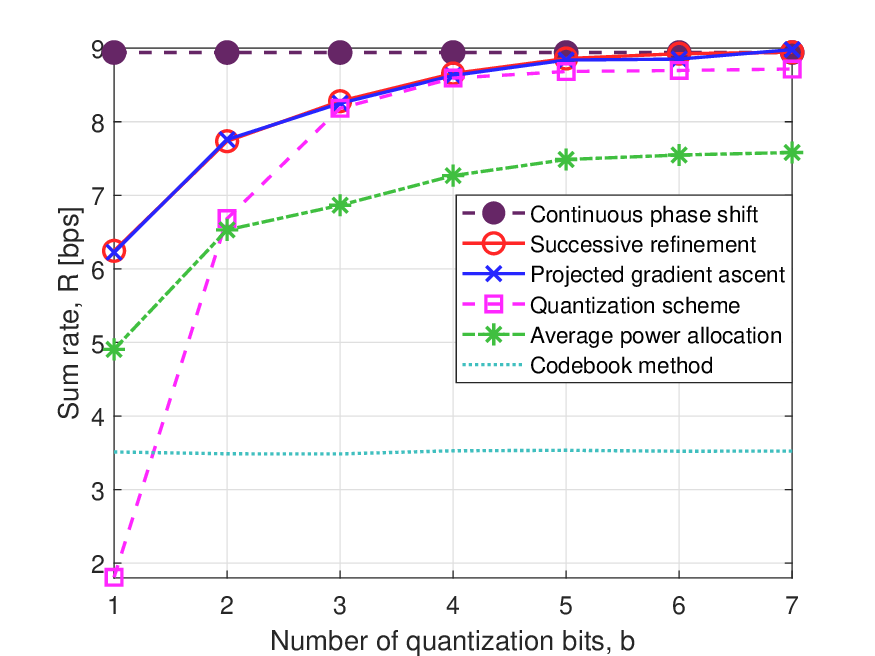}
\caption{Sum rate $R$ versus the number of quantization bits $b$ ($N=49$, $L=7$, $M=K=4$, $P_{\text{T}}=10$ dBm).}
\label{fig_5}
\end{figure}
\begin{figure}[!t]
\centering
\includegraphics[width=8.5cm]{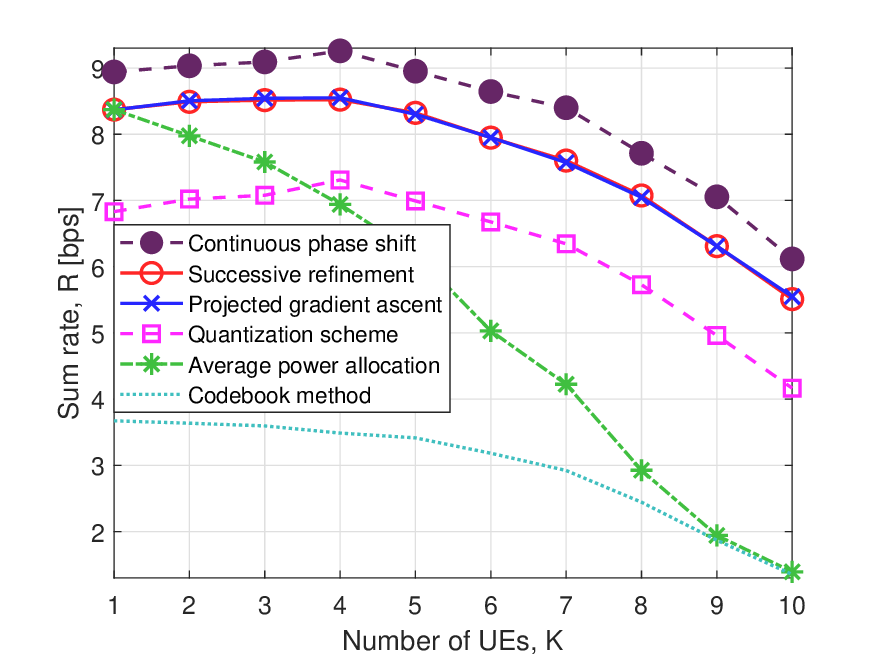}
\caption{Sum rate $R$ versus the number of UEs $K$ ($N=49$, $b=2$, $L = 7$, $P_{\text{T}}=10$ dBm).}\vspace{-0.5cm}
\label{fig_6}
\end{figure}
Fig. \ref{fig_4} depicts the sum rate $R$ versus the number of meta-atoms $N$ on each metasurface layer, using the same system parameters as in Fig. \ref{fig_3}. It is evident that all benchmark schemes attain a performance improvement as $N$ increases. Note that the SIM employing discrete phase shifts with $b = 2$ exhibits a rate penalty of only $0.3$ bps compared to its continuous counterpart for all considered settings. Furthermore, Fig. \ref{fig_4} reveals that the performance gap between the modified iterative water-filling solution and the average power allocation solution gradually diminishes as $N$ increases, which is due to the improved SINR resulting from the larger array aperture of the SIM. In addition, the proposed successive refinement algorithm consistently outperforms the simple and often utilized quantization scheme, particularly as the number of meta-atoms increases. For example, a significant performance gain of 125\% rate increase is attained when $N = 100$. In contrast, the codebook method exhibits only marginal gains as $N$ increases, which is primarily due to the exponential growth of the search space for legitimate phase shift vectors with respect to $N$.

\begin{figure}[!t]
\centering
\includegraphics[width=8.5cm]{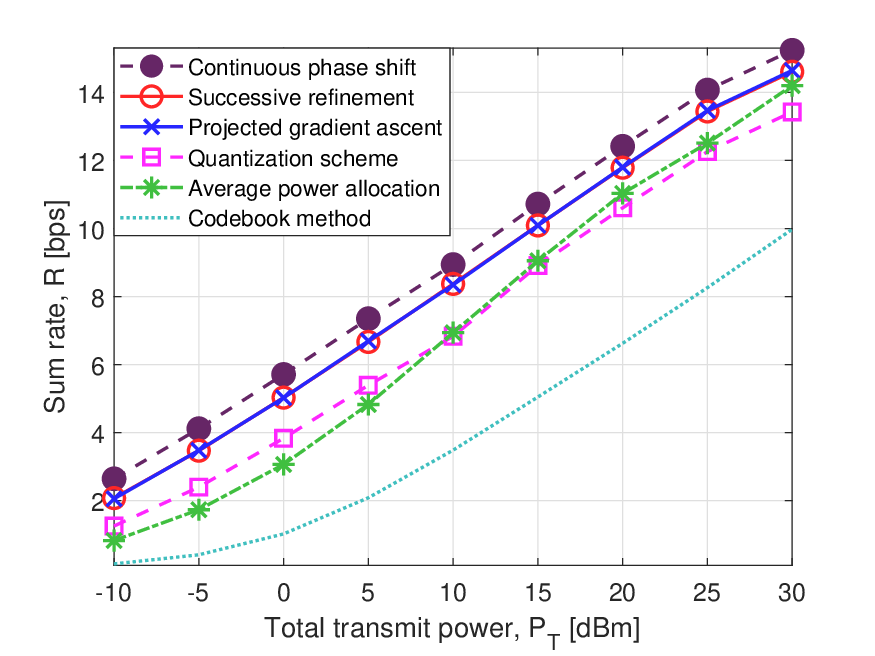}
\caption{Sum rate $R$ versus the transmit power $P_{\text{T}}$ ($N=49$, $b=2$, $L = 7$, $M=K=4$).}
\label{fig_7}
\end{figure}
\begin{figure}[!t]
\centering
\includegraphics[width=8.5cm]{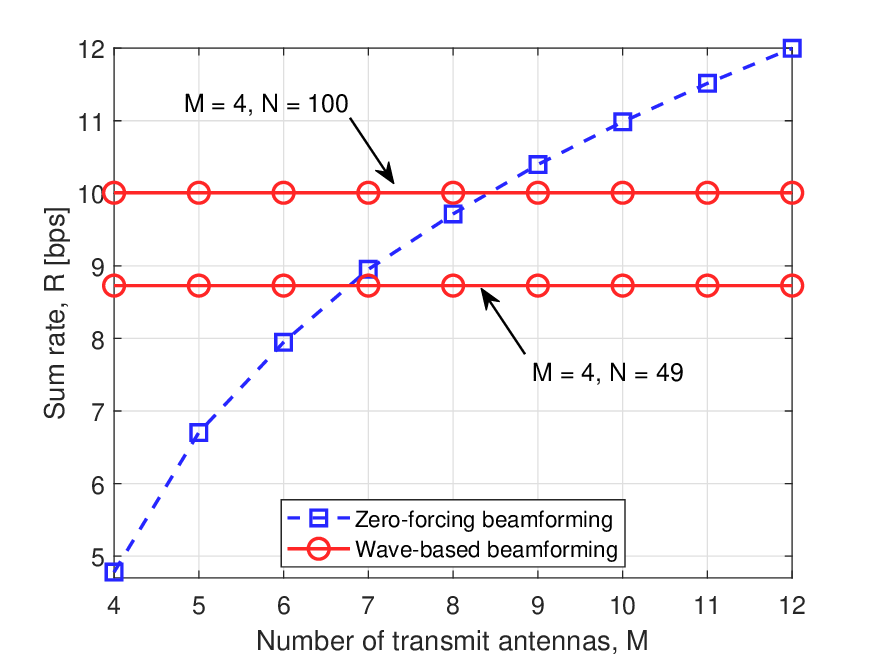}
\caption{Sum rate comparison between the wave-based beamforming and the conventional ZF beamforming.}\vspace{-0.5cm}
\label{fig_8}
\end{figure}
Fig. \ref{fig_5} shows the sum rate versus the number of bits used for quantizing each phase shift, i.e., $b$, where we set $L = 7$ while keeping the other parameters consistent with Fig. \ref{fig_3}. As expected, the proposed successive refinement and gradient ascent scheme as well as the quantization method exhibit a gradual improvement in sum rate as $b$ increases, resulting in performance close to that achieved with continuous phase shifts when $b \geq 6$. Specifically, the quantization scheme suffers from a slight rate loss of $0.2$ bps. Moreover, the sum rate employing the average power allocation solution increases with $b$, but consistently incurs a rate loss of $1.2$ bps compared to the iterative water-filling solution. Additionally, the rate performance of the codebook method remains unaffected by the number of quantization bits due to the limited search space.

In Fig. \ref{fig_6}, we plot the achievable sum rate versus the number of UEs $K$ by setting $b = 2$. All the other parameters remain the same as those in Fig. \ref{fig_5}. As $K$ grows, the sum rate with optimized power allocation and wave-based beamforming initially increases with $K$ but decreases when $K$ exceeds a certain threshold. This is due to the fact that the sum rate of an SIM-assisted multiuser MISO system depends on two factors: the \emph{spatial multiplexing gain} brought by an increased number of UEs as well as the \emph{residual interference level} after applying the wave-based beamforming. For a large number of UEs, it becomes more challenging for the SIM to mitigate the inter-user interference. Additionally, compared to the conventional water-filling algorithm for interference-free parallel channels, the iterative water-filling power allocation method used for interference channels is more prone to be trapped in a local optimum. As a result, the sum rate is expected to decrease when the number of UEs exceeds a certain value. Nevertheless, in practical systems, one could employ user activation and diverse multiple access strategies to ensure that all the selected users are served with the desired QoS requirements. In the specific setup with $L = 7$ and $N = 49$, the maximum number of UEs for which the SIM can effectively mitigate the multiuser interference is $K = 4$. Increasing the number of meta-atoms $N$ could potentially support a greater number of UEs. Furthermore, the quantization method results in a $1.5$ bps rate loss under all setups considered in Fig. \ref{fig_6}, while the average power allocation and codebook method fail to adequately suppress the multiuser interference, resulting in performance degradation as the number of UEs increases. For example, the iterative water-filling algorithm achieves a remarkable 400\% rate improvement over the average benchmark scheme when the number of UEs is $K = 10$.

Fig. \ref{fig_7} illustrates the sum rate of various schemes versus the transmit power $P_{\text{T}}$ by setting $K = 4$. As expected, the sum rate of all benchmark schemes improves as the transmit power $P_{\text{T}}$ increases. In addition, the proposed successive refinement and projected gradient ascent algorithm outperform both the quantization and codebook methods, exhibiting almost identical performance to continuous phase shifts. Notably, the superiority of the proposed algorithm over the codebook method becomes more pronounced as the transmit power increases, resulting in a rate improvement of $4.5$ bps in the moderate and high transmit power regions. Moreover, in the lower power region, the average power allocation suffers from performance erosion as compared to the modified iterative water-filling algorithm. The performance gap, however, becomes negligible as the transmit power grows, which is due to the fact that the water-filling level is much lower than the power allocated for each UE, rendering the impact of the power allocation negligible.

In Fig. \ref{fig_8}, we compare the sum rate of the SIM-assisted multiuser MISO system and its counterpart without an SIM. In the conventional multiuser MISO downlink system, the zero-forcing (ZF) beamforming technique is employed to eliminate the multiuser interference. To ensure a fair comparison, we set $d_{k}=\sqrt{ H_{\text{BS}}^{2}+\left [ d _{\text{UE}}\left ( k-1 \right ) \right ]^{2}}$ to characterize the path loss associated with each UE. The power allocation among the $K$ UEs is obtained by applying the conventional water-filling algorithm \cite{TCOM_20202_An_Low}. We observe that the conventional MISO system requires $M = 7$ transmit antennas to outperform the SIM-assisted systems with $M = 4$ antennas when considering $N = 49$. When considering the same number of transmit antennas, e.g., $M = 4$, the SIM-assisted system achieves about 200\% sum rate improvement compared to the conventional MISO system, thanks to the utilization of larger-aperture metasurfaces. As mentioned already, thanks to the wave-based beamforming enabled by the SIM, the digital precoding is completely avoided at the BS, and each antenna is equipped only with low-resolution ADCs/DACs, which substantially reduces the hardware cost and energy consumption. Nevertheless, the SIM optimization process adds some computational complexity. A more precise performance evaluation between the conventional digital precoding and wave-based precoding schemes requires further study through system-level simulations.

\begin{figure}[!t]
\centering
\includegraphics[width=8.5cm]{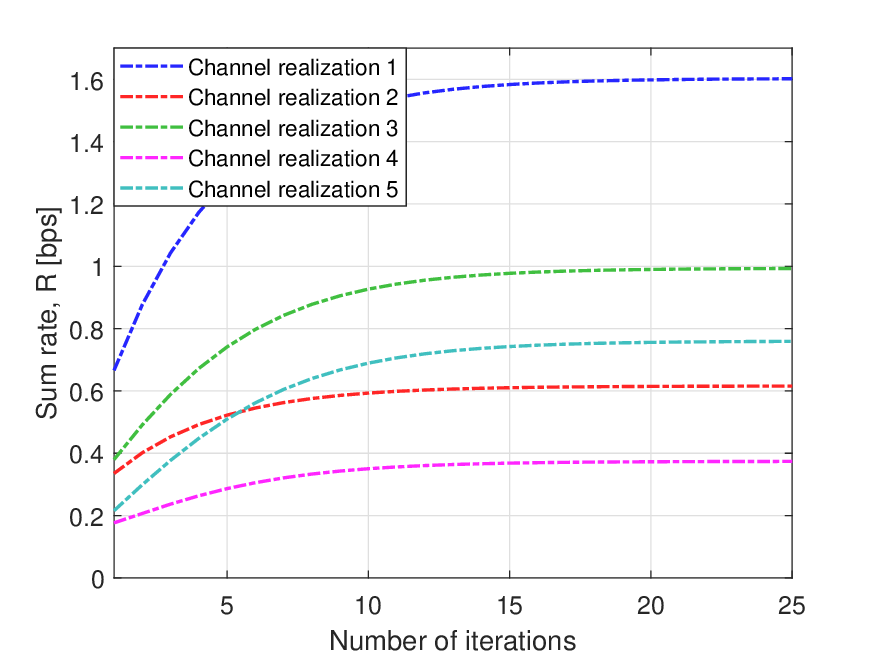}
\caption{Convergence behavior of the iterative water-filling algorithm.}
\label{fig_9}
\end{figure}
\begin{figure}[!t]
\centering
\includegraphics[width=8.5cm]{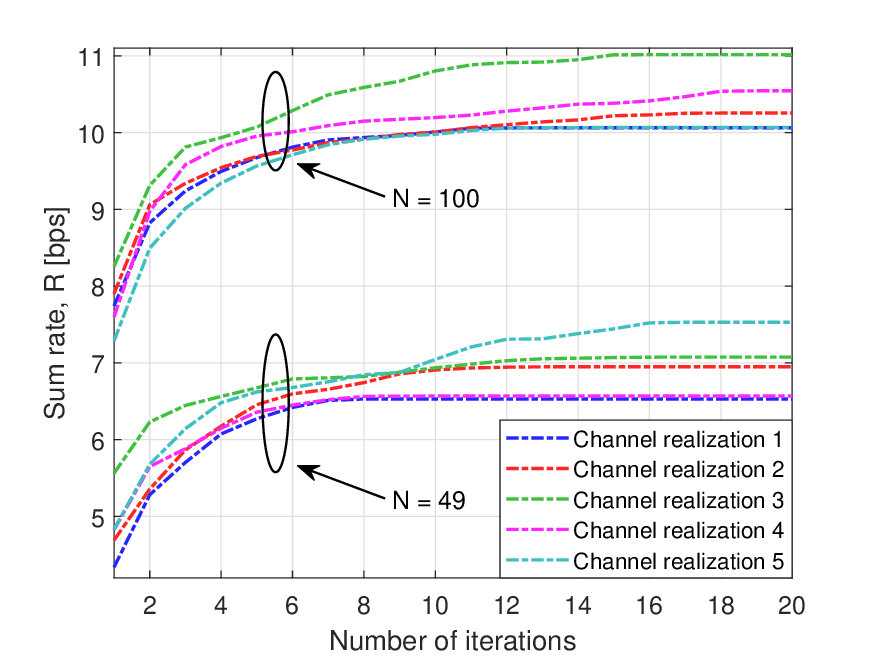}
\caption{Convergence behavior of the successive refinement algorithm.}\vspace{-0.5cm}
\label{fig_10}
\end{figure}
\begin{figure}[!t]
\centering
\includegraphics[width=8.5cm]{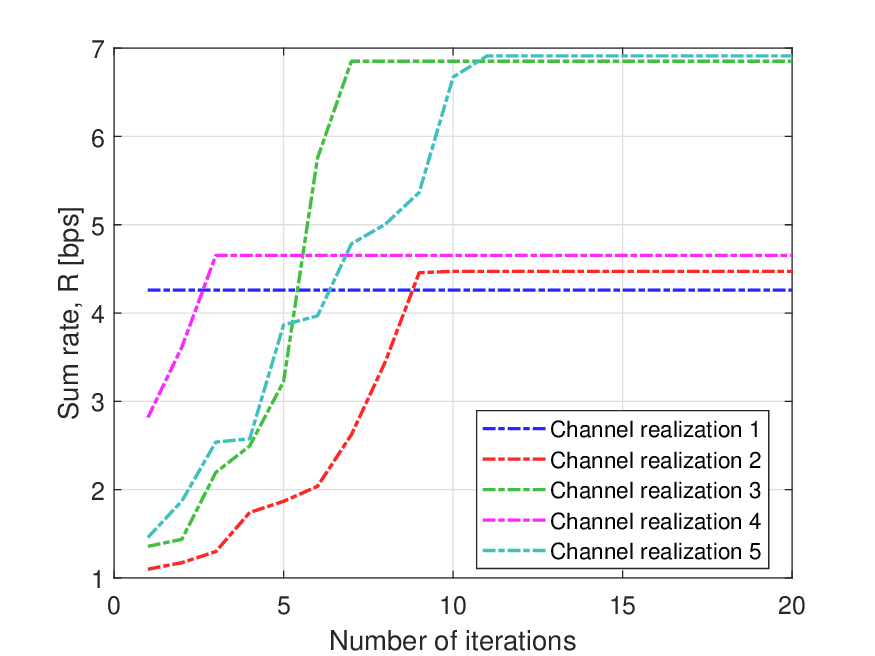}
\caption{Convergence behavior of the projected gradient ascent algorithm.}
\label{fig_11}
\end{figure}
\begin{figure}[!t]
\centering
\includegraphics[width=8.5cm]{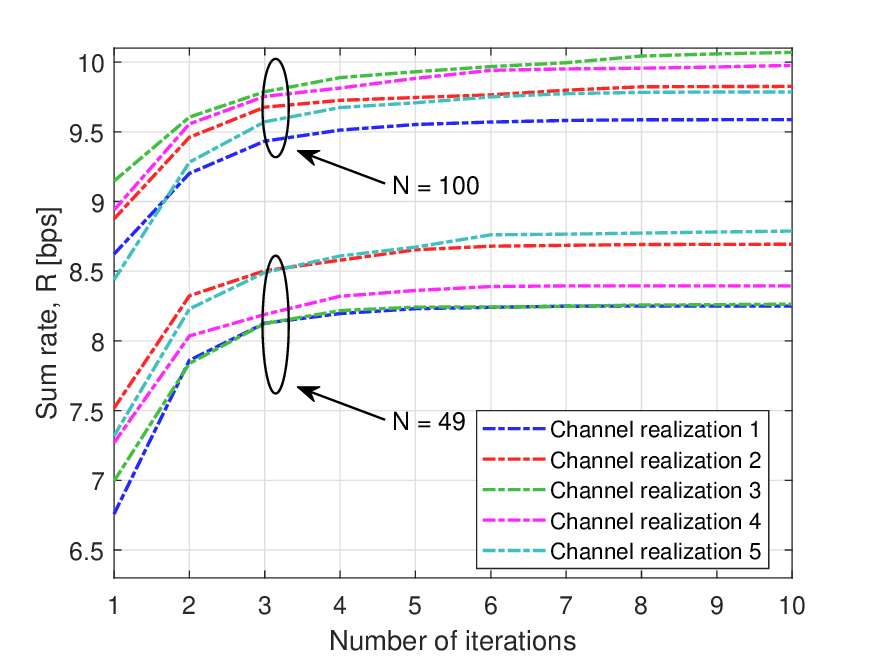}
\caption{Convergence behavior of the AO algorithm.}\vspace{-0.5cm}
\label{fig_12}
\end{figure}
\subsection{Convergence of the Proposed Algorithms}
Next, we verify the convergence of the algorithms proposed in Section \ref{sec3-2}. Fig. \ref{fig_9} shows the convergence behavior of the modified iterative water-filling algorithms, while keeping the transmit power fixed at $P_{\text{T}} = 10$ dBm. Five independent channel realizations with random phase shifts are considered. Although the sum rate varies significantly due to the random phase shifts, we observe that the iterative water-filling algorithm converges to its maximum, which is consistent with the theoretical analysis in \cite{TIT_2005_Jindal_Sum}. Furthermore, the modified iterative water-filling achieves its maximum within $20$ iterations, indicating that $I_{\text{IWF}} \leq 20$ in terms of computational complexity in Section \ref{sec4_3}.

In Fig. \ref{fig_10}, we demonstrate the convergence behavior of the proposed successive refinement algorithm and the projected gradient ascent method, taking into account the average power allocation. We see the rapid convergence of the successive refinement algorithm for two cases: $N = 49$ and $N=100$. In both setups, the univariate optimization method guarantees that the objective function value is non-decreasing and converges within $I_{\text{SR}} \leq 17$ iterations. Nonetheless, at each iteration, increasing the number of meta-atoms requires optimizing more phase shifts, resulting in increased complexity.

Moreover, Fig. \ref{fig_11} analyzes the convergence behavior of the projected gradient ascent algorithm, considering $N = 49$. It is observed that the projected gradient ascent method converges more rapidly compared to the successive refinement method, resulting in $I_{\text{GA}} \leq 10$ under the considered setups. Nevertheless, the quantization process at each iteration may cause the algorithm to be trapped in a local optimum. Thus, under some cases the resultant sum rate in Fig. \ref{fig_11} is slightly lower than that in Fig. \ref{fig_10}, but this can be improved by employing multiple sets of initial phase shifts.

\begin{figure}[!t]
\centering
\includegraphics[width=8.5cm]{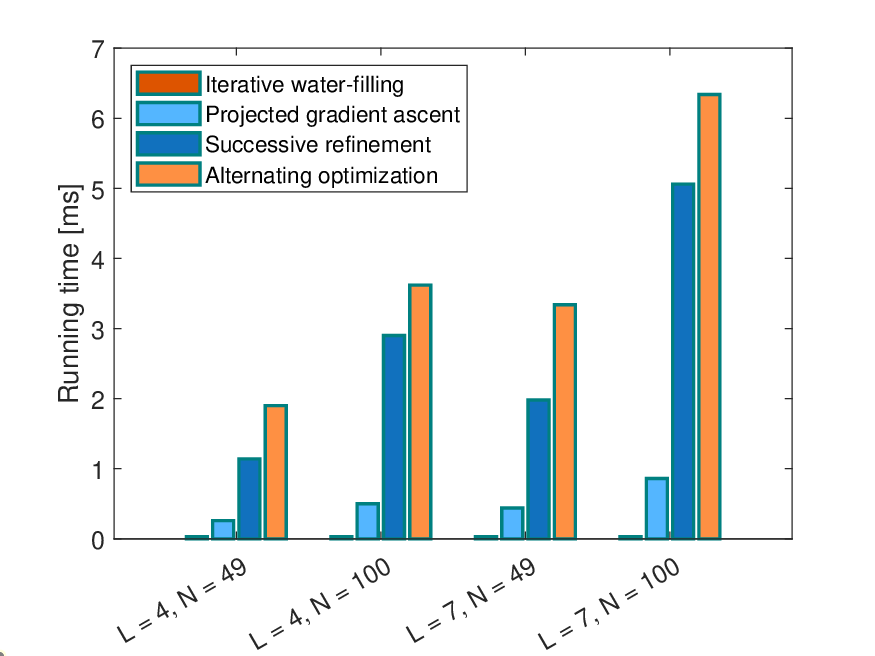}
\caption{Simulation time of the proposed algorithms.}\vspace{-0.5cm}
\label{fig_13}
\end{figure}
Fig. \ref{fig_12} examines the convergence of the AO algorithm, considering the case studies $N = 49$ and $N =100$. In each case, $10$ independent channel realizations are considered. Notably, the outer loop of the AO algorithm converges very fast, reaching its maximum within $7$ iterations under all setups. Therefore, we have $I_{\text{AO}} \leq 7$ in terms of complexity evaluation. Additionally, it is observed that increasing the number of meta-atoms leads to a greater number of iterations required for convergence, which is due to the larger amount of optimization variables and the expanded search space.

In Fig. \ref{fig_13}, finally, we evaluate the run time of the proposed algorithm by utilizing MATLAB R2023b on an AMD Ryzen 9 5950X processor. As shown in the figure, four different SIM setups are considered, and all results are obtained by averaging $1,000$ experiments. It is noted that the iterative water-filling algorithm has the lowest computational complexity, which is independent of the SIM parameters. By contrast, the complexity of the two phase shift optimization methods increases linearly with both $L$ and $N$. Compared to the successive refinement method that tunes the phase shifts one-by-one, the projected gradient ascent method updates the phase shifts simultaneously and thus leads to a reduced simulation time.

\section{Conclusions}\label{sec7}
In this paper, a novel SIM design was proposed for wave-based beamforming in the downlink of a multiuser MISO system. A joint transmit power allocation and phase shift optimization problem was formulated to maximize the sum rate, subject to a transmit power budget and discrete-valued phase shifts. To tackle this problem, an AO algorithm was developed to decompose the original problem into two subproblems. The power allocation subproblem was solved by applying a modified iterative water-filling algorithm, while the phase shifts of the SIM were optimized through the projected gradient ascent algorithm and the successive refinement algorithm. Furthermore, we provided insights into the potential benefits of SIM-based implementations compared to conventional digital beamforming. Extensive simulation results demonstrated that the proposed joint optimization schemes achieve significant performance gains compared to existing benchmarks.

Finally, the major conclusions drawn in this paper can be are summarized as follows. \emph{First}, increasing the number of metasurface layers can lead to an increase in the sum rate, thanks to the interference cancellation ability of the SIM. Simulation results showcased that an SIM performs best with seven layers. \emph{Second}, when the number of quantizing bits is greater than four, the performance of an SIM with discretely tuned meta-atoms is almost identical to the continuous case. \emph{Third}, compared to conventional MISO systems, an SIM-based system provided a 200\% increase in the sum rate. In a nutshell, the SIM technology represents a paradigm shift from conventional digital signal processing to wave-based computation, paving the way toward energy-efficient wireless networks.

\begin{appendices}
\section{Proof of Proposition 1}\label{A1}
The gradient of $R$ with respect to $\theta _{n}^{l}$ can be expressed as
\begin{align}\label{eq29}
\frac{\partial R}{\partial \theta _{n}^{l}}=\log_{2}e\sum_{k=1}^{K}\frac{1}{1+\gamma _{k}}\frac{\partial \gamma _{k}}{\partial \theta _{n}^{l}},\quad \forall n \in \mathcal{N},\quad \forall l \in \mathcal{L}.
\end{align}

Based on the standard quotient rule derivative, $\frac{\partial \gamma _{k}}{\partial \theta _{n}^{l}},\ \forall k \in \mathcal{K}$ in \eqref{eq29} is given by
\begin{align}\label{eq30}
\frac{\partial \gamma _{k}}{\partial \theta _{n}^{l}} &= \frac{p _{k}}{\sum\limits_{{k}'\neq k}^{K}\left | \mathbf{h}_{k}^{\text{H}}\mathbf{G}\mathbf{w}_{{k}'}^{1} \right |^{2}p _{{k}'} + \sigma _{k}^{2}}\frac{\partial \left | \mathbf{h}_{k}^{\text{H}}\mathbf{G}\mathbf{w}_{k}^{1} \right |^{2}}{\partial \theta _{n}^{l}}\notag\\
 &-\frac{\left | \mathbf{h}_{k}^{\text{H}}\mathbf{G}\mathbf{w}_{k}^{1} \right |^{2} p _{k}}{\left ( \sum\limits_{{k}'\neq k}^{K}\left | \mathbf{h}_{k}^{\text{H}}\mathbf{G}\mathbf{w}_{{k}'}^{1} \right |^{2}p _{{k}'} + \sigma _{k}^{2} \right )^{2}}\sum\limits_{{k}'\neq k}^{K}p _{{k}'}\frac{\partial \left | \mathbf{h}_{k}^{\text{H}}\mathbf{G}\mathbf{w}_{{k}'}^{1} \right |^{2}}{\partial \theta _{n}^{l}}.
\end{align}

By utilizing \eqref{eq30}, the partial derivative $\frac{\partial R}{\partial \theta _{n}^{l}}$ in \eqref{eq29} can be further simplified to
\begin{align}\label{eq31}
\frac{\partial R}{\partial \theta _{n}^{l}}=&\log_{2}e\sum_{k=1}^{K}\delta _{k}\times\notag\\
 &\left ( p _{k}\frac{\partial \left | \mathbf{h}_{k}^{\text{H}}\mathbf{G}\mathbf{w}_{k}^{1} \right |^{2}}{\partial \theta _{n}^{l}}-\gamma _{k}\sum\limits_{{k}'\neq k}^{K}p _{{k}'}\frac{\partial \left | \mathbf{h}_{k}^{\text{H}}\mathbf{G}\mathbf{w}_{{k}'}^{1} \right |^{2}}{\partial \theta _{n}^{l}} \right ),
\end{align}
where $\delta _{k}$ is defined in \eqref{eqq19}.

The key challenge in \eqref{eq31} lies in determining the partial derivative of $\left | \mathbf{h}_{k}^{\text{H}}\mathbf{G}\mathbf{w}_{{k}'}^{1} \right |^{2}$ with respect to $\theta _{n}^{l}$. Note that $\mathbf{h}_{k}^{\text{H}}\mathbf{G}\mathbf{w}_{{k}'}^{1}$ is linear with respect to $e^{j\theta _{n}^{l}}$. Therefore, for any given pair $\left ( k,k' \right )\in \mathcal{K}^{2}$, we have
\begin{align}\label{eq32}
\frac{\partial \left | \mathbf{h}_{k}^{\text{H}}\mathbf{G}\mathbf{w}_{{k}'}^{1} \right |^{2}}{\partial \theta _{n}^{l}}&=\frac{\partial \left | \sum\limits_{n=1}^{N}e^{j\theta _{n}^{l}}\mathbf{h}_{k}^{\text{H}}\mathbf{v}_{n}^{l}\left ( \mathbf{u}_{n}^{l} \right )^{\text{H}}\mathbf{w}_{{k}'}^{1} \right |^{2}}{\partial \theta _{n}^{l}}\notag\\
&=\frac{\partial \Re\left [ \left ( e^{j\theta _{n}^{l}}\mathbf{h}_{k}^{\text{H}}\mathbf{v}_{n}^{l}\left ( \mathbf{u}_{n}^{l} \right )^{\text{H}}\mathbf{w}_{{k}'}^{1} \right )\left ( \mathbf{h}_{k}^{\text{H}}\mathbf{G}\mathbf{w}_{{k}'}^{1} \right )^{\text{H}} \right ]}{\partial \theta _{n}^{l}}\notag\\
&=\Im\left [ \left ( e^{j\theta _{n}^{l}}\mathbf{h}_{k}^{\text{H}}\mathbf{v}_{n}^{l}\left ( \mathbf{u}_{n}^{l} \right )^{\text{H}}\mathbf{w}_{{k}'}^{1} \right )^{\text{H}}\left ( \mathbf{h}_{k}^{\text{H}}\mathbf{G}\mathbf{w}_{{k}'}^{1} \right ) \right ]\notag\\
&=\eta _{n,k,{k}'}^{l},
\end{align}
where $\eta _{n,k,{k}'}^{l}$, $\left ( \mathbf{u}_{n}^{l} \right )^{\text{H}}$, and $\mathbf{v}_{n}^{l}$ are defined in \eqref{eqq20}, \eqref{eqq21}, and \eqref{eqq22}, respectively.

Upon substituting \eqref{eq32} into \eqref{eq31}, the proof is completed. $\hfill \blacksquare$
\end{appendices}
\bibliography{ref}
\bibliographystyle{IEEEtran}

\end{document}